\documentclass[%
preprint, 
bibnotes,
amsmath,amssymb,
aps, physrev, prx
]{revtex4-2}

\usepackage{graphicx}
\usepackage{dcolumn}
\usepackage{bm}
\usepackage[dvipsnames]{xcolor}

\begin{document}

\preprint{APS/123-QED}

\title{\textbf{Dissipative anomalies of stresses in soft amorphous solids: footprints of density singularities} 
}%

\author{Anier Hernández-García}
 \affiliation{Physics Department, University of Colorado at Boulder}
 \email{Contact author: anier.hernandezgarcia@colorado.edu}



\date{\today}

\begin{abstract}
In soft amorphous solids, localized irreversible (plastic) stress dissipation occurs as a response to external forcings \cite{Barbot_PhysRevE.97.033001}. A crucial question is whether we can identify structural properties linked to a region's propensity to undergo a plastic stress drop when thermal effects are negligible \cite{Nicolas_RevModPhys.90.045006, PhysRevMaterials.4.113609}. To address this question, I follow a theoretical framework provided by Onsager's ideal turbulence theory, representing a non-perturbative application of the renormalization group scale-invariance principle \cite{eyink2018review, Eyink_2024}. First, I analyze the zero temperature limit for the fine-grained balance equation for the stress tensor corresponding to instanton realizations. I show that irreversible stress drops can occur if the density gradients diverge. I then derive a balance relation for the coarse-grained instantaneous stress tensor with arbitrary regularization scale $\ell$. From the latter, I obtain an expression for the local inter-scale stress flux in terms of moments of the density increments. This expression is analogous to Kolmogorov's $\frac{4}{5}th\;law$ for incompressible fluid turbulence, but valid for individual realizations. By assuming that the density field is Besov regular, I determine the scaling of the stress flux with $\ell$. From this scaling, I show that distributional solutions of the noiseless Dean (NDE) equation can sustain stress dissipation due to a non-equilibrium inter-scale stress flux if the scaling exponents of the density structure functions are below a critical threshold. The athermal limit of fine-grained and coarse-grained descriptions must describe the same phenomenology, \textit{i.e.} the existence of stress dissipation must be independent of any regularization of the dynamics. Using this principle, I analyze the limit $\ell\rightarrow 0$ and argue that flow realizations of athermal disordered systems correspond to ultraviolet fixed-point solutions of the coarse-grained NDE equations with sufficiently low Besov regularity.

\section{Introduction}

Amorphous materials, including colloidal and molecular glasses, granular matter, and foams,  exhibit highly nonlinear rheological properties. These materials flow when a stress threshold is exceeded, while at lower loads, their response is solid-like \cite{Larson_Rheology}. Numerous industrial applications,  ranging from the design of new materials to structure development, rely upon the elaboration of predictive theories to describe the deformation and flow of such materials. A critical starting point of such theories is the identification of appropriate structural variables to describe these systems. This topic presents significant challenges, as noted in \cite{Parisi_Glasses}; we cannot describe the dynamics of disordered (glassy) systems using conventional plane waves, and the absence of a "small" parameter renders most attempts at perturbative treatments ineffective. Thus, studying of the physical mechanisms and mathematical properties underlying the dissipation of stresses in disordered solids are of paramount importance from both applied and fundamental points of view. 

Nevertheless, despite differences in typical disorder scales and mechanical stiffness, the phenomenology of this vast class of materials exhibits remarkable universal features when subjected to external loading \cite{Nicolas_RevModPhys.90.045006, Parisi_PNAS2017}. These materials respond in a quasi-linear elastic regime for relatively small imposed strain or forces. However, once the induced stresses exceed a local threshold, denoted as the local yield stress $\sigma_c$ \cite{Sylvain_PhysRevLett.117.045501, Barbot_PhysRevE.97.033001}, they undergo small-scale irreversible (plastic) deformations, which manifest as local irreversible particle rearrangements (LIPR). These events exhibit a broad distribution of sizes and shapes \cite{Rodney_2011, Barbot_PhysRevE.97.033001}. A key empirical observation is that following such LIPRs, small-scale regions experience irreversible stress drops, which also lead to a decrease in the total or spatial average of stresses across the material. Consequently, a large-scale anisotropic redistribution of stresses occurs in the remaining parts of the material (see, for instance, Fig. 1b in \cite{Barbot_PhysRevE.97.033001}).

A systematic characterization of the elementary mechanisms that lead to irreversible stress drops during small-scale plastic relaxation events is essential to develop predictive theories of mechanical responses in amorphous materials. These theories should effectively connect the local and global rheology of these systems.  To our knowledge, no rigorous analytical work has adequately addressed these issues from first principles.

 In contrast, theories of crystal plasticity have been applied with some success. It is well-established that crystallographic defects and dislocations act as sources of plastic relaxation events (\cite{Kleman_Lavrentovich}, Chapter 8). By studying the dynamics of these localized topological defects, researchers have been able to develop rigorous constitutive equations for crystal plasticity. Similarly, observations made in an athermal glassy model system \cite{Barbot_PhysRevE.97.033001} led the authors to suggest that, analogously to crystal plasticity, the existence of plastic deformations via discrete units encoded in the material structure cannot be ruled out. These discrete units are thought to preexist in the material before loading. Their findings suggest that the plastic deformations of an amorphous solid, at least in the small deformation regime, can be understood as a sequence of activated discrete shear transformation zones with weak slip orientations. Thus, it seems natural to pose the question:\textit{"Could similar static structural defects be identified in the absence of a regular structure? The question has been vivid to the present day, so that it is fair to say that, at least, they are much more elusive than in crystals."} \cite{Nicolas_RevModPhys.90.045006}.  In this work, I address this question. 

 Many studies have shown that the plasticity of soft amorphous solids may be linked to structural defects \cite{Nicolas_RevModPhys.90.045006, PhysRevMaterials.4.113609}. As discussed in these works, researchers have introduced various indicators encompassing structural, thermodynamic, and mechanical information about materials. Examples include energetically favorable regions, linear and non-linear vibrational properties, local thermal energy, and local shear modulus. These measures aim to characterize the softness of different regions within the material. While these indicators generally show good correlations with plastic activity at low strain, they do not effectively elucidate the underlying mechanisms that lead to plastic events \cite{Baggioli2023, PhysRevMaterials.4.113609}. Conversely, in coarse-grained phenomenological models of plasticity, defects are assumed but without a precise definition \cite{Baggioli2023}. 


 In this paper, I develop a first-principles theory based on the hypothesis that plastic rearrangements result from thermal dissipative anomalies of stresses. In the limit of vanishing temperatures, I argue that distributional solutions of the noiseless Dean equation describe the system's dynamics. Additionally, I show that not only dislocations but also density configurations with regions of low Besov spatial regularity can act as sources of small-scale plastic deformations, thereby broadening the definition of "defects" in disordered media (section III.B).

  After completing the work presented in this paper, I became aware of the remarkable findings in \cite{Wu2023}. This research directly correlated topological defects in the static structure of glasses prior to deformation with their subsequent plastic behavior. The authors of \cite{Wu2023} demonstrated that the spatial distribution of vibrational eigenvectors exhibits singular behavior. They found that in amorphous systems subjected to quasistatic shear, there is a strong correlation between plastic events and regions with a high density of frustrated interfaces featuring a saddle shape. 

The theory I develop here extends beyond the non-equilibrium scenarios considered in \cite{Wu2023}, as it is not limited to quasistatic shear loading. It remains valid even when non-smooth, small-scale external forces affect the material. Additionally, it does not require the assumption of homogeneous small-scale elasticity.

The structure of the article is organized as follows: In section (II),  I use the Martin-Siggia-Rose/Janssen-DeDominicis path integral representation of the stochastic equation for the density field to obtain the equations for the action minimizers, known as instantons. In section (III),  I derive a balance relation for the stress tensor corresponding to the instanton realizations. I then investigate the implications of having singular density gradients for stress conservation laws. To achieve this, I smear the stress balance equation and analyze its distributional limit. Next, in subsection (III.A), I derive an effective low-wavenumber equation for the density field in the athermal limit by coarse-graining the stochastic Dean equation. From this equation, I examine a balance relation for the large-scale mechanical stresses at an arbitrary space resolution, denoted as $\ell$ in subsection (III.B). Following this, I express the sources/sinks of stresses in terms of the density increments.   The latter result is an analog of Kolmogorov's $\frac{4}{5}th\;law$ for incompressible fluid turbulence, but valid for individual realizations of amorphous solids. By considering the limit $T\rightarrow0$ for both the instanton description and the coarse-grained equations with regularization scale $\ell$, and using the principle of renormalization group invariance,  I provide a precise characterization of the singularities of the density field required to have sources of plastic deformations. At the end of that section, I  present concluding remarks and suggest perspectives.

 

\section{Path Integral formulation and instanton equations.}

To investigate the rheology of soft amorphous solids, let us chose a model system consisting of overdamped Brownian particles interacting via a pairwise potential $V$. In such a model, the governing equations for the dynamics of the particles are

\begin{equation}
\frac{d \mathbf{x}_{\alpha}}{dt}= - \mu \sum_{\alpha<\nu}\frac{\partial V(\mathbf{x}_{\alpha}(t)-\mathbf{x}_\nu(t))}{\partial \mathbf{x}_{\alpha}(t)} + \mu \mathbf{f}^e(\mathbf{x}_{\alpha}(t),t)+\mathbf{\xi}_{\alpha}(t),
\label{Langevinsystem}
\end{equation}

in which the vector noise $\mathbf{\xi}_{\alpha}(t)$ acting on particle $\alpha$  has probability

\begin{equation}
P[\mathbf{\xi}_{\alpha}]\propto exp\left(-\frac{1}{2}\int d\tau  \frac{\xi_{\alpha}^2(\tau) }{2 k_b T \mu}\right),
\label{probnoise}
\end{equation}
and it is not correlated with the noise acting on particle $\nu$. The external force is denoted by $\mathbf{f}^e(\mathbf{x}_{\alpha}(t),t)$. It is convenient to cast the system of Eqs. \ref{Langevinsystem} in terms of the dimensionless variables 

\begin{equation}
\mathbf{x}^{*}=\mathbf{x}/\ell_o\;\;\;\;\;\;\;V^{*}=V/V_0 \;\;\;\; t^{*}= \frac{\mu V_0}{\ell_o^2}t\;\;\;\; \mathbf{f}^{e*} =  \frac{\ell_0}{V_0}\mathbf{f}^{e}.
\label{dimensionless variables}
\end{equation} 

In terms of these variables we obtain

\begin{equation}
\frac{d \mathbf{x}^{*}_{\alpha}}{dt}= -\sum_{{\alpha}<\nu}\frac{\partial V^{*}(\mathbf{x}_{\alpha}^{*}(t^{*})-\mathbf{x}_\nu^{*}(t^{*}))}{\partial \mathbf{x}_{\alpha}^{*}(t^{*})} + \mathbf{f}^e(\mathbf{x}_{\alpha}^{*}(t^{*}),t^{*})+\mathbf{\xi}_{\alpha}^{*}(t^{*}),
\label{Langevinsystemdim}
\end{equation}
with
 \begin{equation}
P[\mathbf{\xi}_{\alpha}^{*}]\propto exp\left(-\frac{1}{2}\int d\tau  \frac{\xi_{\alpha}^2(\tau) }{2 T_v}\right),
\end{equation}
in which $T_v= k_b T/V_o$. It is convenient to chose the spatial scale $\ell_0$ as the characteristic interaction length of the pairwise potential $V$. To simplify the notation, we will omit the superscript symbol $"^{*}"$  and a dimensionless formulation must be understood, unless indicated otherwise.

 As shown in \cite{Dean_1996}, with the Ito discretization convention the system of Eqs.(\ref{Langevinsystem}) can be reformulated as an \textit{exact} and \textit{closed} Langevin equation for the density operator $\rho(\mathbf{x},t)=\sum_{i}\delta(\mathbf{x}-\mathbf{x}_i (t))$

 \begin{equation}
   \frac{\partial \rho(\mathbf{x},t)}{\partial t}= \nabla\cdot \left(\vec{\eta}(\mathbf{x},t)\sqrt{\rho(\mathbf{x},t)}\right) + \nabla\cdot\left(\int d^d x^{'} \rho(\mathbf{x},t)\nabla V(\mathbf{x}-\mathbf{x}^{'})\rho(\mathbf{x}^{'},t)\right) + T_v\nabla^2 \rho - \nabla\cdot (\rho(\mathbf{x},t) \mathbf{f}^e(\mathbf{x},t))
   \label{deanequation},
\end{equation} 
in which the  noise $\mathbf{\eta(\mathbf{x},t)}$ has probability

\begin{equation}
P[\vec{\eta}(\mathbf{x},t)]\propto exp\left(-\frac{1}{2}\int d^d \varsigma\int d\tau  \frac{\eta^2(\varsigma,\tau) }{2 T_v }\right).
\label{probdensitynoise}
\end{equation}

As emphasized in the works (\cite{Marconi_1999},\cite{Donev_dfth_hyd}), Eq. \ref{deanequation} is an informative nontrivial rewriting of Eq. \ref{dimensionless variables} that serves as a starting point for coarse-graining and describes the dynamics at different scales. Indeed, in \cite{Kruger_stresses}, from the Dean equation (\ref{deanequation}) the authors identified the divergence of the instantaneous stress tensor 
$\sigma(\mathbf{x},t)$ as 

\begin{equation}
   \nabla\cdot \mathbb{\sigma}(\mathbf{x},t) = -T_v\nabla \rho(\mathbf{x},t) -\rho(\mathbf{x},t)\nabla \int d x^{'} V(\mathbf{x}-\mathbf{x}^{'})\rho(\mathbf{x}^{'},t). 
   \label{divergencestress}
   \end{equation} 
   
   The behavior of the stress tensor components, arising from the interaction term that includes the potential $V$, depends on the joint statistics of the two-point quadratic density function. To assess the statistical properties of these quantities, it is helpful to first cast the underlying dynamical Eq. (\ref{deanequation}) into its functional representation. In this work, the Martin-Siggia-Rose/Janssen-De Dominicis-Peliti (MSRJDP) response field formalism is adopted \cite{MSR_PhysRevA.8.423,Peliti_Dominic_PhysRevB.18.353,DeDominicis,Janssen1976}. Within this formalism, the dynamical partition function is expressed as follows:


 \begin{equation}
 \mathcal{Z} = \int D\rho\int D\hat{\rho}e^{i \mathcal{S[\rho,\hat{\rho}}]},
 \label{MSRJD}
\end{equation}  
in which the action functional $\mathcal{S[\rho,\hat{\rho}}]$ is determined by

\begin{equation}
\mathcal{S[\rho,\hat{\rho}}]= \int d^d \varsigma\int d\tau \hat{\rho}\left(\frac{\partial \rho}{\partial t}-\mathcal{L}(\rho)\right)+i T_v\int d^d \varsigma\int d\tau \rho(\nabla\hat{\rho})^2,
\label{MSRaction}
\end{equation}
where $\mathcal{L}(\rho) = \nabla\cdot\left(\int d^d x^{'} \nabla^{'} V(\mathbf{x}^{'})\rho(\varsigma)\rho(\varsigma-x^{'})\right) + T_v\nabla^2 \rho - \nabla\cdot (\rho \mathbf{f}^e) $ 


As a first step, the functional integral in Eq. (\ref{MSRaction}) is evaluated in the stationary path approximation. The corresponding equations for the action minimizers paths (instantons) are derived by the variation of $\mathcal{S[\rho,\hat{\rho}}]$ with respect to $\rho$ and $\hat{\rho}$. This variation yields the saddle-point equations: 

\begin{equation}
\frac{\partial \rho}{\partial t}-\mathcal{L}(\rho)=-T_v\nabla\cdot(\rho\nabla\hat{\rho}),
\label{instrho}
\end{equation}
and
\begin{equation}
\frac{\partial \hat{\rho}}{\partial t}+\hat{\rho}\frac{ \delta L[\rho]}{\delta\rho}=-T_v(\nabla\hat{\rho})^2.
\label{instrhohat}
\end{equation}

Eq. (\ref{instrho}) must be supplemented with the initial condition $\rho(x,t_0)= \rho_0(x)$ at $t_0$, while for the response field $\hat{\rho}$ the boundary condition is implied at the remote future $\hat{\rho}$, since from Eq. (\ref{instrhohat}) it propagates in the negative direction of time. 

Since the stress tensor is a functional of the density field, solving the system of Eqs. (\ref{instrho}) and (\ref{instrhohat}) allows us to completely characterize its behavior corresponding to instanton realizations.



\section{Dissipative anomalies of mechanical stresses. }

  Let us  compute a pointwise balance relation for the internal stress tensor at a nonzero temperature $T_v$. For the sake of clarity of the analysis let us denote the \textit{smooth} solution of Eq. (\ref{instrho}) at a finite temperature as $\rho^s(\mathbf{x})$. As in \cite{Kruger_stresses},  let us assume that the interaction potential satisfies  reflection symmetry $\nabla V(\mathbf{x})=-\nabla V(-\mathbf{x})$. As discussed in the latter work this consideration  allows to invert the divergence in Eq. (\ref{divergencestress}). This procedure gives for the instantaneous internal stress tensor the expression (17) in \cite{Kruger_stresses}, which is included here for completeness 
  
   \begin{equation}
   \mathbf{\sigma}(\mathbf{x},t)=- T \rho^s(\mathbf{x},t)\mathfrak{I} + \frac{1}{2}\int_{0}^{1} d \lambda\int d x^{'} \mathbf{x}^{'}\nabla^{'} V(\mathbf{x}^{'})
 \rho^s(\mathbf{x}-\lambda \mathbf{x}^{'},t )\rho^s(\mathbf{x}+(1-\lambda)\mathbf{x}^{'},t),
   \label{stresstensor}
   \end{equation}
   where $\mathfrak{I}$ denotes the identity matrix.
  
To simplify our analysis, let's first focus on the instanton associated with cases where the response field $\hat{\rho}$ identically vanishes. By taking the partial derivative with respect to time of Eq. \eqref{stresstensor}, and applying Eq. \eqref{instrho} for the time derivative of the density field, we obtain the following balance equation for the stress tensor, valid at each point:
  
  \begin{equation}
  \frac{\partial \sigma_{ij}^s}{\partial t}+ \partial_{x_k}J^s_{kij}= \mathcal{Q}_{ij}^{thermal}+\mathcal{W}_{ij}^{ni}+\mathcal{F}_{ij}^e,
  \label{stressbalancetemp}
  \end{equation}
  where
  \begin{equation}
  J_{kij}^s=T_v\left(\partial_{x_l}\sigma_{kl}-\rho^s(\mathbf{x},t)f^e_k(\mathbf{x},t)\right)\delta_{ij}+ J^s_{int,kij},
  \label{spacetransport}
  \end{equation}
  with
  \begin{equation}
  J_{int,kij}^s=-\frac{1}{2}\int_{0}^{1} d\lambda\int d^d y y_j\nabla_{y_i}V(\mathbf{y})\left\lbrace T\nabla_k(\rho^{'s}\rho^s) + \rho^{'s}\rho^s(f^{ni}_k+f^{'ni}_k-(f^{'e}_k+f^e_k))\right\rbrace,
  \end{equation}

  \begin{equation}
  \mathcal{F}_{ij}^e=-\frac{1}{2}\int_{0}^{1} d\lambda\int d^d y y_j\nabla_{y_i}V(\mathbf{y})\left(\nabla\rho^s\cdot\rho^{'s}\mathbf{f}^e+\nabla\rho^{'s}\cdot\rho\mathbf{f}^{'e}\right),
  \label{externalinput}
  \end{equation}
  
  \begin{equation}
  \mathcal{Q}_{ij}^{thermal}=-T_v\int_{0}^{1} d\lambda\int d^d y y_j\nabla_{y_i}V(\mathbf{y})\nabla\rho^s\cdot\nabla\rho^{'s}
  \label{thermaldissipation}
  \end{equation}
  and 
  \begin{equation}
  \mathcal{W}_{ij}^{ni}=-\frac{1}{2}\int_{0}^{1} d\lambda\int d^d y y_i\nabla_{y_j}V(\mathbf{y})\left(\rho^{'s}\mathbf{f}^{'ni}\cdot\nabla\rho^s+\rho^s\mathbf{f}^{ni}\cdot\nabla\rho^{'s}\right),
  \label{productionint}
  \end{equation}
  with the notations 
  
  $$\rho=\rho(\mathbf{x}-\lambda \mathbf{y},t)\;\;\;\;\rho^{'}=\rho(\mathbf{x}+(1-\lambda) \mathbf{y},t)\;\;\;\;\mathbf{f}^{ni}=\mathbf{f}^{ni}(\mathbf{x}-\lambda \mathbf{y},t)\;\;\;\;\mathbf{f}^{'ni}=\mathbf{f}^{ni}(\mathbf{x}+(1-\lambda) \mathbf{y},t),$$
  
  in which $\mathbf{f}^{ni}(\mathbf{x},t)\equiv  \int d^d x^{'} \nabla V(\mathbf{x}^{'})\rho(\mathbf{x}-\mathbf{x}^{'},t)$ is the total internal force density acting on the point $\mathbf{x}$.


The balance Eq. (\ref{stressbalancetemp}) applies to \textit{individual} instanton realizations. I derived this equation assuming infinite or periodic systems, where the interaction potential obeys reflection symmetry. The term $\mathcal{F}_{ij}^e$ represents the external input of stresses. Both the interparticle interactions and thermal diffusive effects will contribute to the time derivative of the total stress tensor through the terms $\mathcal{W}_{ij}^{ni}$ and $\mathcal{Q}_{ij}^{thermal}$, respectively. The term $J_{kij}$ is the \textit{k-th} component of the vector flux for the \textit{ij} component of $\sigma$.

Note the similarity between $\mathcal{Q}_{ij}^{thermal}$ and the viscous kinetic energy dissipation rate found in both incompressible and compressible fluid flows (\cite{Landau_Lifshitz_FluidMechanics}, chapter II section 16.). In the case of soft amorphous solids, this term involves a non-local multiplication of the density gradients. The term (\ref{thermaldissipation}) will vanish for smooth density fields when $T_v\rightarrow 0$. However, this does not necessarily hold if the density gradients diverge. We observe that the thermal term contains the higher-order derivatives in the instanton equations, suggesting that a singular behavior may emerge as $T_v\rightarrow 0$. This situation is reminiscent of the dissipative anomaly that turbulent flows exhibit in the limit $Re\rightarrow \infty$ \cite{Eyink_2024, BENZI20231}. Several coarse-grained modeling approaches, such as elastoplastic models, have postulated the existence of non-analyticities in the strain and stress fields \cite{Nicolas_RevModPhys.90.045006, Baggioli2023}. As highlighted in the Introduction, recent research has shown that singularities in the topology of the vibrational eigenmodes in two-dimensional glasses directly relate to their plastic behavior in systems undergoing athermal quasi-static simple shear \cite{Wu2023}. In this work, I will investigate the implications of having singular density gradients for stress conservation laws.

\subsection{Distributional limit of the fined-grained stress balance equation. }

 Let's now investigate the distributional limit of the fine-grained stress balance Eq. (\ref{stressbalancetemp}) when $T_v\rightarrow 0$. Smearing Eq. (\ref{stressbalancetemp}) with a function $\phi(\mathbf{x},t)\in D\left((0,T)\times \Omega, \mathbb{R}^{3}\right)$, compactly supported $supp \; \phi \subset (0,T)\times \Omega$ we obtain 

 \begin{align}
   \int d^dx\left(\sigma_{ij}^s(\mathbf{x},T)\phi(\mathbf{x},T)-\sigma_{ij}^s(\mathbf{x},0)\phi(\mathbf{x},0)\right)  -\int_0^T dt \int d^{d}x \sigma_{ij}(\mathbf{x},t)\frac{\partial \phi}{\partial t} = \nonumber \\
   -\frac{1}{2}\int d^dx^{'}\int_0^1 d\lambda\int_0^Tdt\int d^d\xi x_i^{'}\partial_{x_j^{'}}V(\mathbf{x}^{'})\rho^s(\xi,t)f_l^{ni}(\xi,t)\partial_{\xi_l}\left[\rho^s(\xi+\mathbf{x}^{'},t)\left(\phi(\xi+\lambda\mathbf{x}^{'})+\phi(\xi+(1-\lambda)\mathbf{x}^{'}\right)\right] \nonumber \\ 
    -\frac{1}{2}\int d^dx^{'}\int_0^1 d\lambda\int_0^Tdt\int d^d\xi x_i^{'}\partial_{x_j^{'}}V(\mathbf{x}^{'})\rho^s(\xi,t)f_l^{e}(\xi,t)\partial_{\xi_l}\left[\rho^s(\xi+\mathbf{x}^{'},t)\left(\phi(\xi+\lambda\mathbf{x}^{'})+\phi(\xi+(1-\lambda)\mathbf{x}^{'}\right)\right] \nonumber \\
    - \frac{T_v}{2} \int d^dx^{'}\int_0^1 d\lambda\int_0^Tdt\int d^d\xi x_i^{'}\partial_{x_j^{'}}V(\mathbf{x}^{'})\partial_{\xi_l}\rho^s(\xi,t)\partial_{\xi_l}\left[\rho^s(\xi+\mathbf{x}^{'},t)\left(\phi(\xi+\lambda\mathbf{x}^{'})+\phi(\xi+(1-\lambda)\mathbf{x}^{'}\right)\right],
    \label{finegraineddistr}
 \end{align}
 where $\rho^s$ denotes solutions of the instanton equations with finite $T_v$. In this work I assume strong convergence of the density to a field $\rho$ when $T_v\rightarrow 0$, $$\rho^s\rightarrow \rho \;\; in  \;\;L^2((0,T),L^2_{loc}(\Omega)).$$

Under this assumption, we observe that the integral in the first line of Eq. \ref{finegraineddistr} converges in the limit as $T_v \rightarrow 0 $. Additionally, if we assume that the integral of the second derivatives of the interparticle potential exists, we can prove through integration by parts that the integrals in the second and third lines of Eq. \ref{finegraineddistr} also converge in the athermal limit. This implies that the distributional limit of $ \mathcal{Q}^s_{ij} $ exists,


\begin{equation}
    \mathcal{Q}_{ij}^{thermal}=\mathcal{D}- \lim_{T_v\rightarrow 0}  \mathcal{Q}^s_{ij}.
\end{equation}

We have the following balance equation,

\begin{align}
\mathcal{D}- \lim_{T_v\rightarrow 0} \left\lbrace\int d^dx\left(\sigma_{kl}^s(\mathbf{x},T)\phi(\mathbf{x},T)-\sigma_{kl}^s(\mathbf{x},0)\phi(\mathbf{x},0)\right)  -\int_0^T dt \int d^{d}x \sigma^s_{kl}(\mathbf{x},t)\frac{\partial \phi}{\partial t}\right\rbrace= \nonumber \\ \mathcal{F}^{ni}_{kl}+ \mathcal{F}^{e}_{kl}+\mathcal{Q}_{kl},
\label{tvdistributionallimit}
\end{align}

in which $\mathcal{F}^{ni}_{kl}$ and $\mathcal{F}^{e}_{kl}$ denote the distributional limits of the second and third lines of Eq. \ref{finegraineddistr}, respectively.

 
 \subsection{Large-scale effective  equations. }

 The thermal dissipation term in Eq. \ref{thermaldissipation} has a structure similar to the viscous kinetic energy dissipation rate observed in both incompressible and compressible turbulence \cite{Monin_Yaglom}. Its contribution to the stress balance approaches zero in the limit $ T_v \rightarrow 0 $ for smooth density fields. However, this may not hold true in the presence of structural singularities, such as crystallographic dislocations. In these cases, the derivatives of the fields in Eqs. \ref{instrho} and \ref{instrhohat} should be interpreted in a generalized distributional sense.
 

 
   Thus, to remove the small scale ultraviolet divergences and investigate low wavenumber phenomena  let us study the evolution of the coarse-grained field given by
 
  \begin{equation}
\overline{\rho}_{\ell}(\mathbf{x},t)=\int d^3r G_{\ell}(\mathbf{r})\rho(\mathbf{x}+\mathbf{r},t),
\label{coarsegrainedensity}
\end{equation}
 in which $G_{\ell}(\mathbf{r})\equiv \ell^{-d}G(\mathbf{r}/\ell)$. The kernel $G(\mathbf{r})$ must satisfy the following general conditions 
 
 $$G(\mathbf{r})\geq 0,$$
 $$\int d^3r G(\mathbf{r})=1,$$
 $$\int d^3r \mathbf{r}G(\mathbf{r})=0, $$ and
$$\int d^3r |\mathbf{r}|^2G(\mathbf{r})=1.$$

Also, it will be required that $G(\mathbf{r})$ be smooth and rapidly decaying in space. Formally, it is  demanded that $G(\mathbf{r})$ belongs to the space of infinitely-differentiable functions with compact support. 

Coarse-graining the stochastic Dean equation \ref{deanequation} yields
  
   \begin{equation}
   \frac{\partial \overline{\rho}_{\ell}}{\partial t}(\mathbf{x},t)= \nabla\cdot \overline{\left(\vec{\eta}(\mathbf{x},t)\sqrt{\rho(\mathbf{x},t)}\right)}+\nabla\cdot \overline{\left[ \rho(\mathbf{x}) \nabla \int d x^{'} V(\mathbf{x}-\mathbf{x}^{'})\rho(\mathbf{x}^{'}) \right]}_{\ell}+ T_v \nabla^2\overline{\rho}_{\ell}- \nabla\cdot \overline{(\rho(\mathbf{x},t) \mathbf{f}^e(\mathbf{x},t))}_{\ell}.
   \label{largescaledeanequation}
\end{equation}

Intuitively, one expects that for the dynamics of sufficiently large-scale modes, the thermal diffusive effects will not have a direct impact. The range of scales for which this is valid should be larger as the temperature decreases and will eventually be valid for all scales $\ell$ in the limit $T_v\rightarrow 0$. Formally, this intuitive reasoning means that in the limit $T_v\rightarrow 0$, the limiting field $\rho$  must satisfy
\begin{equation}
 \frac{\partial \rho(\mathbf{x},t)}{\partial t}= \nabla\cdot\left(\int d^d x^{'} \rho(\mathbf{x},t)\nabla V(\mathbf{x}-\mathbf{x}^{'})\rho(\mathbf{x}^{'},t)\right) - \nabla\cdot (\rho(\mathbf{x},t) \mathbf{f}^e(\mathbf{x},t))
   \label{athermaldeanequation}
\end{equation}
in a distributional sense. This is equivalent to saying that the limiting field $\rho$ is a coarse-grained solution of the NDE (\cite{EyinkDrivas2017}, see Proposition 1). Hence, the following equation must be satisfied pointwise

 \begin{equation}
   \frac{\partial \overline{\rho}_{\ell}}{\partial t}(\mathbf{x},t)= \nabla\cdot \overline{\left[ \rho(\mathbf{x}) \nabla \int d x^{'} V(\mathbf{x}-\mathbf{x}^{'})\rho(\mathbf{x}^{'}) \right]}_{\ell}- \nabla\cdot \overline{(\rho(\mathbf{x},t) \mathbf{f}^e(\mathbf{x},t))}_{\ell}.
   \label{lsadeanequation}
\end{equation}

To demonstrate this, one must show that the diffusive and noise terms in Eq. \ref{largescaledeanequation} vanish pointwise, while the other terms converge to those in the coarse-grained NDE as $T_v \rightarrow 0$.
.


Let us estimate the coarse-grained diffusive term $T_v\nabla^2\overline{\rho}_{\ell}$. Integration by parts leads to

\begin{equation}
 T_v\nabla^2\overline{\rho}_{\ell}=T_v\ell^{-2}\int d^d r \left(\nabla_r^2 G\right)_{\ell}(\mathbf{r})\delta\rho(\mathbf{r};\mathbf{x}),
\label{cgthermalterm}
\end{equation}
in which $\delta\rho(\mathbf{r};\mathbf{x})=\rho(\mathbf{x}+\mathbf{r})-\rho(\mathbf{x})$, \cite{Course-Eyink} \href{https://www.ams.jhu.edu/~eyink/Turbulence/}{Turbulence course, G. Eyink}. From this identity it is seen that $T_v \nabla^2\overline{\rho}_{\ell}=\mathcal{O}(T_v\ell^{-2}\delta\rho(\ell))$, where $\delta\rho(\ell)$ denotes a typical density increment over the scale $\ell$. 

This estimate indicates that for a fixed value of $\ell$, the diffusive term approaches zero as $T_v$ decreases. Conversely, for a fixed temperature, this term vanishes in the limit $\ell\rightarrow\infty$: it is an irrelevant term in the technical sense of the RG scale invariance principle. To evaluate the noise term, it is also useful to apply integration by parts, which yields the following result:

\begin{equation}
\nabla\cdot \overline{\left(\vec{\eta}(\mathbf{x},t)\sqrt{\rho(\mathbf{x},t)}\right)}=-\frac{\sqrt{2 T_v}}{\ell}\int d^d r \left(\nabla G\right)_{\ell}(\mathbf{r})\delta(\sqrt{\rho}\vec{\eta^{'}}(\mathbf{r};\mathbf{x}))
\label{cgnoiseterm},
\end{equation}
where $\vec{\eta}\equiv\sqrt{2 T_v}\vec{\eta^{'}}$.

From the equality \ref{cgnoiseterm} and the definition of the noise $\vec{\eta}^{'}$ it is seen that at any fixed $\ell > 0$, this term also vanishes in the athermal limit. Hence, for $T_v \ll 1$, the \textit{direct} effects of both the thermal diffusive and noise terms are negligible at any fixed $\ell > 0$. It follows from these arguments that as $T_v$ vanishes, the athermal limiting density field satisfies for all $\ell$ the coarse-grained equation
\begin{equation}
   \frac{\partial \overline{\rho}_{\ell}}{\partial t}(\mathbf{x},t)= \nabla\cdot \overline{\left[ \rho(\mathbf{x}) \nabla \int d x^{'} V(\mathbf{x}-\mathbf{x}^{'})\rho(\mathbf{x}^{'}) \right]}_{\ell}- \nabla\cdot \overline{(\rho(\mathbf{x},t) \mathbf{f}^e(\mathbf{x},t))}_{\ell}.
   \label{lsadeanequation}
\end{equation}
As discussed in (\cite{EyinkDrivas2017}, see Proposition 1 of section 2 and theorem 2) this statement is equivalent to say that the limiting density field is a distributional solution of the NDE.

   \subsection{Dynamics of the large-scale stress tensor for coarse-grained solutions of the athermal Dean equation.}
 In this section I will derive a balance equation in the athermal limit for the resolved or large-scale non-ideal component of the stress tensor
  
     \begin{equation}
   \underline{\mathbf{\sigma}}^{ni}_{\ell}(\mathbf{x},t)= \frac{1}{2}\int d x^{'} \mathbf{x}^{'}\nabla V(\mathbf{x}^{'})\int_{0}^{1} d \lambda
 \rho_{\ell}(\mathbf{x}-\lambda \mathbf{x}^{'} )\rho_{\ell}(\mathbf{x}+(1-\lambda)\mathbf{x}^{'}),
   \label{largescalestresstensor}
   \end{equation}
   in which for $\underline{\mathbf{\sigma}}_{\ell}$ I have used the same shorthand notation as in \cite{EyinkDrivas2017,PhysRevX.8.011022}. I emphasize that the coarse-graining scale $\ell$ is arbitrary.


    The time derivative of $ \underline{\mathbf{\sigma}}^{ni}_{\ell}(\mathbf{x},t)$ can be expressed as

       \begin{equation}
  \frac{\partial \underline{\mathbf{\sigma}}^{ni}_{\ell}(\mathbf{x},t) }{\partial t} = \frac{1}{2}\int d x^{'} \mathbf{x}^{'}\nabla V(\mathbf{x}^{'})\int_{0}^{1} d \lambda
\left\lbrace \frac{\partial \rho_{\ell}(\mathbf{x}-\lambda \mathbf{x}^{'} )}{\partial t}\rho_{\ell}(\mathbf{x}+(1-\lambda)\mathbf{x}^{'})+ \rho_{\ell}(\mathbf{x}-\lambda \mathbf{x}^{'} )\frac{\partial \rho_{\ell}(\mathbf{x}-\lambda \mathbf{x}^{'} )}{\partial t}\right\rbrace.
   \label{largescalestresstensortimeder}
   \end{equation}
   
 The coarse-grained NDE Eq. (\ref{largescaledeanequation}) can be equivalently written as
 
 \begin{equation}
   \frac{\partial \overline{\rho}_{\ell}}{\partial t}(\mathbf{x},t)= \nabla\cdot \left[\overline{\rho}_{\ell}(\mathbf{x}) \left(\overline{\mathbf{f}^{ni}_{\ell}}(\mathbf{x},t)-\overline{\mathbf{f}^{e}_{\ell}}(\mathbf{x},t)\right)+\tau_{\ell}(\rho, \mathbf{f}^{ni}) -\tau_{\ell}(\rho, \mathbf{f}^{e})\right] \equiv-\nabla\cdot\mathbf{j}_{\ell}(\mathbf{x},t), 
   \label{largescaledeanequationcum}
\end{equation}
where $\tau_{\ell}(f, g)=\overline{(fg)}_{\ell}-\overline{f}_{\ell} \overline{g}_{\ell}$ is the second order cumulant of the "variables" $f$ and $g$. In the last part of Eq. \ref{largescaledeanequation} the term  $$\mathbf{j}_{\ell}(\mathbf{x},t)= \int d^d r G_{\ell}(\mathbf{r})\delta(\mathbf{x}+\mathbf{r}-\mathbf{x_i}(t))\frac{d \mathbf{x_i}(t)}{d t}$$ represent the large scale or resolved instantaneous current. A key identity is \cite{Constantin1994, Eyink1995}

\begin{equation}
\tau_{\ell}(f, g)= \int d^dr G_{\ell}(\mathbf{r})\delta f(\mathbf{r};\mathbf{x})\delta g(\mathbf{r};\mathbf{x})-\int d^dr G_{\ell}(\mathbf{r})\delta f(\mathbf{r};\mathbf{x})\int d^dr^{'} G_{\ell}(\mathbf{r}^{'})\delta g(\mathbf{r}^{'};\mathbf{x})\equiv \tau_{\ell}(\delta f, \delta g).
\label{identitycumulantsstress}
\end{equation}

The quantity $\tau_{\ell}(\delta\rho, \delta F_i)$ constitutes a "small-scale or unresolved" force due to particle interactions.


 Inserting Eq. \ref{largescaledeanequationcum} in Eq. \ref{largescalestresstensortimeder} and using the properties of the divergence operator one obtains

  \begin{equation}
 \partial_t \underline{\sigma}_{\ell} (\mathbf{x})+\nabla\cdot \overline{\mathcal{J}}_{\ell}(\mathbf{x})=\mathcal{W}_{\ell}(\mathbf{x})+\mathcal{Q}^{flux}_{\ell}(\mathbf{x}),
  \label{resolvedstresstensor}
\end{equation}
 in which 
 \begin{equation}
 \overline{\mathcal{J}}_{\ell}(\mathbf{x})=\frac{1}{2}\int d^dx^{'}\int_{0}^{1}d\lambda\mathbf{x}^{'}\nabla V(\mathbf{x}^{'})\overline{\mathbf{J}}_{\ell}(\mathbf{x}-\lambda\mathbf{x}^{'} ,\mathbf{x}+(1-\lambda)\mathbf{x}^{'}),
 \label{fluxlsstress}
 \end{equation}
 is the spatial transport (flux) vector of large-scale stresses,
 
 \begin{equation}
 \mathcal{W}_{\ell}(\mathbf{x})=\frac{1}{2}\int d^dx^{'}\int_{0}^{1}d\lambda\mathbf{x}^{'}\nabla V(\mathbf{x}^{'})W_{\ell}(\mathbf{x}-\lambda\mathbf{x}^{'} ,\mathbf{x}+(1-\lambda)\mathbf{x}^{'}), 
 \label{pressureworklsstress}
\end{equation}  
and
\begin{equation}
\mathcal{Q}^{flux}_{\ell}(\mathbf{x})=\frac{1}{2}\int d^dx^{'}\int_{0}^{1}d\lambda\mathbf{x}^{'}\nabla V(\mathbf{x}^{'})Q_{\ell}^{flux}(\mathbf{x}-\lambda\mathbf{x}^{'} ,\mathbf{x}+(1-\lambda)\mathbf{x}^{'}). 
\label{cascadestress}
\end{equation} 

are the sources/sinks of the non-ideal component of stresses resulting from inter-particle forces. In Eqs. \ref{fluxlsstress}, \ref{pressureworklsstress} and \ref{cascadestress} I have defined the quantities 

\begin{equation}
\overline{\mathbf{J}}_{\ell}(\mathbf{x}_1,\mathbf{x}_2)=-\left[\overline{\rho}_{\ell}(\mathbf{x}_2)\nabla\cdot\underline{\sigma}_{\ell}(\mathbf{x}_1) + \overline{\rho}_{\ell}(\mathbf{x}_1)\nabla\cdot\underline{\sigma}_{\ell}(\mathbf{x}_2) +\overline{\rho}_{\ell}(\mathbf{x}_2)\tau_{\ell}(\rho, \mathbf{F})(\mathbf{x}_1)+\overline{\rho}_{\ell}(\mathbf{x}_1)\tau_{\ell}(\rho, \mathbf{F})(\mathbf{x}_2)\right],
\label{spatialfluxpointsplitqdf}
\end{equation}

 \begin{equation}
 W_{\ell}(\mathbf{x}_1,\mathbf{x}_2)=-\left[\nabla\overline{\rho}_{\ell}(\mathbf{x}_2)\cdot(\nabla\cdot\underline{\sigma}_{\ell})(\mathbf{x}_1) + \nabla\overline{\rho}_{\ell}(\mathbf{x}_1)\cdot(\nabla\cdot\underline{\sigma}_{\ell})(\mathbf{x}_2)\right],
 \label{pressureworkpointsplitqdf}
 \end{equation}
and 
 \begin{equation}
Q^{flux}_{\ell}(\mathbf{x}_1,\mathbf{x}_2)=-\left[ \nabla\overline{\rho}_{\ell}(\mathbf{x}_2)\cdot\tau_{\ell}(\rho, \mathbf{F})(\mathbf{x}_1)+\nabla\overline{\rho}_{\ell}(\mathbf{x}_1)\cdot\tau_{\ell}(\rho, \mathbf{F})(\mathbf{x}_2)\right]
 \label{interscaleintegrand},
 \end{equation}

 respectively. 

 The system of Eqs. \ref{pressureworkpointsplitqdf}-\ref{interscaleintegrand} provides a scale-by-scale budget of stresses for an arbitrary resolution scale $\ell$. Eqs. \ref{pressureworklsstress} and \ref{pressureworkpointsplitqdf} indicate that $\mathcal{W}_{\ell}(\mathbf{x})$ represents a pure large-scale quantity. In contrast,  Eqs. \ref{cascadestress} and \ref{pressureworkpointsplitqdf} show that $\mathcal{Q}^{flux}_{\ell}(\mathbf{x})$ is a typical "fluxlike" term, i.e., it describes an interaction between the resolved density gradients $\nabla\overline{\rho}_{\ell}$ and the sub-scale forces in the system due to particle interactions $\tau_{\ell}(\rho, \mathbf{F})$.   This term accounts for the scale-to-scale flux of the non-ideal component of stresses. We note that the mathematical structure of $\mathcal{Q}^{flux}_{\ell}(\mathbf{x})$ is similar to the energy cascade terms found in high-Re turbulence. This similarity is expected because $\underline{\mathbf{\sigma}}_{\ell}$ depends quadratically on the density field, and the nonlinearity in Eq. (\ref{largescaledeanequation}) involves the spatial derivatives of a quadratic function of the density. Consequently,  Eqs. \ref{pressureworklsstress}-\ref{pressureworkpointsplitqdf} represent nonlinear mechanisms responsible for the non-conservation of stresses at a given resolution length scale $\ell$.

   The coarse-graining procedure developed in this section eliminates the ultraviolet divergencies of the density field. While the dynamics of the stress tensor may change with the resolution scale $\ell$, a key point is that non-trivial conclusions can be drawn by varying the coarse-graining scale $\ell$ and demanding that any physical phenomenon must be $\ell-$ independent.

 \subsection{Fine-grained stress balance equation. }
 
Exploiting the freedom to vary $\ell$, let us now analyze the limit $\ell\rightarrow 0$ of the balance Eq. \ref{resolvedstresstensor}. For the estimates derived in what follows, we will assume that $\rho$, $V$  and $\nabla V$ $\in L^{\infty}(\Omega)$, which imply that these functions are locally p-integrable $\forall \; p\geq 1$. For any open set $O \subset\subset \Omega$, I will use the notation $\parallel \cdot \parallel_{p,O}$ for the standard $L_p$-norm on the restriction $\rho\mid _O$.  Additionally, we will assume that the support  $supp(G)$ is contained within the Euclidean unit ball for convenience.

 To analyze the limit $\ell\rightarrow 0$ for the system of Eqs. \ref{pressureworklsstress}-\ref{pressureworkpointsplitqdf} I first smear it with a function $ \phi(\mathbf{x},t)\in D\left((0,T)\times \Omega, \mathbb{R}^{3}\right)$, compactly supported $supp \; \phi \subset (0,T)\times \Omega$. This leads to

 \begin{align}
   \int d^dx\left(\underline{\sigma}_{ij,\ell}(\mathbf{x},T)\phi(\mathbf{x},T)-\underline{\sigma}_{ij}^s(\mathbf{x},0)\phi(\mathbf{x},0)\right)  -\int_0^T dt \int d^{d}x \underline{\sigma}_{ij,\ell}(\mathbf{x},t)\frac{\partial \phi}{\partial t} = \nonumber \\
   +\frac{1}{2}\int d^dx^{'}\int_0^1 d\lambda\int_0^Tdt\int d^d\xi x_i^{'}\partial_{x_j^{'}}V(\mathbf{x}^{'})\rho_{\ell}(\xi,t)F_{k,\ell}^{ni}(\xi,t)\partial_{\xi_k}\left[\rho_{\ell}(\xi+\mathbf{x}^{'},t)\left(\phi(\xi+\lambda\mathbf{x}^{'})+\phi(\xi+(1-\lambda)\mathbf{x}^{'}\right)\right] \nonumber \\ 
   +\frac{1}{2}\int d^dx^{'}\int_0^1 d\lambda\int_0^Tdt\int d^d\xi x_i^{'}\partial_{x_j^{'}}V(\mathbf{x}^{'})\tau_{k,\ell}(\delta \rho, \delta \mathbf{F}^{ni})(\xi,t)\partial_{\xi_k}\left[\rho_{\ell}(\xi+\mathbf{x}^{'},t)\left(\phi(\xi+\lambda\mathbf{x}^{'})+\phi(\xi+(1-\lambda)\mathbf{x}^{'}\right)\right] \nonumber \\ 
    -\frac{1}{2}\int d^dx^{'}\int_0^1 d\lambda\int_0^Tdt\int d^d\xi x_i^{'}\partial_{x_j^{'}}V(\mathbf{x}^{'})\rho_{\ell}(\xi,t)f_{k,\ell}^{e}(\xi,t)\partial_{\xi_k}\left[\rho_{\ell}(\xi+\mathbf{x}^{'},t)\left(\phi(\xi+\lambda\mathbf{x}^{'})+\phi(\xi+(1-\lambda)\mathbf{x}^{'}\right)\right] \nonumber \\
    - \frac{1}{2} \int d^dx^{'}\int_0^1 d\lambda\int_0^Tdt\int d^d\xi x_i^{'}\partial_{x_j^{'}}V(\mathbf{x}^{'})\tau_{k,\ell}(\delta \rho, \delta \mathbf{f}^{e})(\xi,t)\partial_{\xi_k}\left[\rho_{\ell}(\xi+\mathbf{x}^{'},t)\left(\phi(\xi+\lambda\mathbf{x}^{'})+\phi(\xi+(1-\lambda)\mathbf{x}^{'}\right)\right].
    \label{regdistributionallimit}
 \end{align}

   Let us denote the term in the second line of Eq. \ref{regdistributionallimit} as $\mathcal{M}_{ij,\ell}$. It is a function of the resolved, large-scale density field and its gradient. Similarly, the fourth line of the Eq. \ref{regdistributionallimit} is a purely large-scale quantity, which we will denote as $\mathcal{N}_{ij,\ell}$. The third term $\mathcal{Q}_{ij,\ell}^{flux}$ is a typical flux-like term that accounts for interactions between resolved and unresolved degrees of freedom. It represents an inter-scale flux of stresses among the different scales due to non-ideal interactions. This term introduces a potential source of stress conservation anomalies for the coarse-grained Dean equation obtained first in the limit of $T_v\rightarrow 0$ and then by taking $\ell\rightarrow 0$. We will discuss below the conditions under which its distributional limit, as $\ell\rightarrow 0$, is non-vanishing. The fifth term $\mathcal{N}^{flux,e}_{ij,\ell}$ is proportional to the work done by the unresolved degrees of freedom of the external force. It is also a fluxlike term. This term generally vanishes in typical experiments where the external load is applied to the material's surface or when the external forcing varies smoothly throughout the material. However, if soft amorphous materials are stirred by non-smooth, small-scale forces—such as those generated by active microorganisms—this term may significantly contribute to the stress transfer among different scales within the material.   

   Let's take the limit $\ell\rightarrow 0$ in Eq. \ref{regdistributionallimit}. Note that $\rho_{\ell}\rightarrow \rho$ when $\ell\rightarrow 0$, see the discussion in \cite{EyinkDrivas2017}. Now consider 
 
\begin{equation}
\Vert \underline{\sigma}_{\ell} - \sigma \Vert_{p,O} = \Vert \frac{1}{2}\int d x^{'} \int_{0}^{1} d \lambda \mathbf{x}^{'}\nabla V(\mathbf{x}^{'})\left[\rho_{\ell}(\mathbf{x}-\lambda \mathbf{x}^{'} )\rho_{\ell}(\mathbf{x}+(1-\lambda)\mathbf{x}^{'})-
 \rho(\mathbf{x}-\lambda \mathbf{x}^{'} )\rho(\mathbf{x}+(1-\lambda)\mathbf{x}^{'}) \right]\Vert_{p,O}.
 \label{stressineq}
\end{equation}

By Minkowski's integral inequality one can obtain 
  \begin{equation} 
  \Vert \underline{\sigma}_{\ell} - \sigma \Vert_{p,O}\leq\frac{1}{2}\int d x^{'} \int_{0}^{1} d \lambda\vert\; \mathbf{x}^{'}\nabla V(\mathbf{x}^{'})\vert \Vert \rho_{\ell}(\mathbf{x}-\lambda \mathbf{x}^{'} )\rho_{\ell}(\mathbf{x}+(1-\lambda)\mathbf{x}^{'})-
 \rho(\mathbf{x}-\lambda \mathbf{x}^{'} )\rho(\mathbf{x}+(1-\lambda)\mathbf{x}^{'})\Vert_{p,O}.
  \label{stressineq2}
\end{equation}
 
 The following inequality also holds
   \begin{equation}
 \Vert \rho_{\ell}\rho^{'}_{\ell}-\rho\rho^{'}\Vert_{p,O}\leq\Vert \overline{(\rho\rho^{'})_{\ell}}-\rho\rho^{'}\Vert_{p,O} +\Vert\overline{(\rho\rho^{'})_{\ell}}-\rho_{\ell}\rho^{'}_{\ell}\Vert_{p,O},
 \label{densityineq}
\end{equation}
in which I used the shorthand notation $\rho=\rho(\mathbf{x}-\lambda \mathbf{x}^{'} )$ and $\rho^{'}=\rho(\mathbf{x}+(1-\lambda)\mathbf{x}^{'})$. If we assume local Besov regularity $\Vert\delta\rho(\mathbf{r})\Vert_{p,O}\leq C |\mathbf{r}|^{\xi_p}$ with the Besov pth-order exponents satisfying $0<\xi_p\leq 1$ then 

\begin{equation}
\lim_{\ell\rightarrow 0} \Vert\overline{(\rho\rho^{'})_{\ell}}-\rho_{\ell}\rho^{'}_{\ell}\Vert_{p,O}=0,
\end{equation}
 where the results from Proposition 3 of \cite{EyinkDrivas2017} have been used. Likewise
 
 \begin{equation}
\lim_{\ell\rightarrow 0} \Vert \overline{(\rho\rho^{'})_{\ell}}-\rho\rho^{'}\Vert_{p,O}=0.
\end{equation}

From the inequalities Eqs. (\ref{densityineq}) and (\ref{stressineq2}) we infer that $\underline{\sigma}_{\ell}$ converges to $\sigma$ strong in $L^p$. This result implies that the distributional limit of the first line in \ref{regdistributionallimit} exists. Consequently, the distributional limit of the sum of the right-hand side in \ref{regdistributionallimit} necessarily exists. Moreover, the fine-grained stress balance Eq. \ref{tvdistributionallimit} in the limit of $T_v\rightarrow 0$ must agree with the coarse-grained stress balance Eq. \ref{regdistributionallimit} in the limit $\ell\rightarrow 0$. That is, the rate of stress variation cannot depend upon the arbitrary scale $\ell$ of spatial coarse-graining. Therefore, the limits $T_v\rightarrow 0$ and $\ell\rightarrow 0$ must commute. By applying this RG scale invariance principle, we obtain  


\begin{equation}
    \mathcal{D}-\lim_{\ell\rightarrow 0}\left(\mathcal{M}_{kq,\ell}^{ni}+\mathcal{N}_{kq,\ell}^e+\mathcal{N}^{flux,e}_{kq,\ell}+\mathcal{Q}_{kq,\ell}^{flux}\right)=\mathcal{F}^{ni}_{kq}+ \mathcal{F}^{e}_{kq}+\mathcal{Q}_{kq}^{thermal}
    \label{RGbalance}
\end{equation}

For the moment, we consider situations in which the system is driven out of equilibrium by the initial conditions and assume that no external forces are acting on the material. Under such assumptions, all the terms in Eq. \ref{RGbalance} that contain information about the external forcing will vanish. We have for the stress conservation anomaly


\begin{equation}
\mathcal{Q}_{kq}^{thermal} = \mathcal{H}_{kq}-\mathcal{Q}_{kq}^{flux}
\label{compactDA}
\end{equation}
with $\mathcal{H}_{kq}= \mathcal{D}-\lim_{\ell\rightarrow 0,T_v\rightarrow 0}\mathcal{M}_{kq,\ell}^{ni}- \mathcal{F}^{ni}_{kq} $ and $\mathcal{Q}_{kq}^{flux}=\mathcal{D}-\lim_{\ell\rightarrow 0} \mathcal{Q}_{kq,\ell}^{flux}$.

Given the definitions of $\mathcal{M}_{kq,\ell}^{ni}$ and $\mathcal{F}^{ni,T_v}_{kq}$ in the second lines of \ref{regdistributionallimit} and \ref{finegraineddistr}, respectively, one might conjecture that $H_{kq}$ vanishes. While the limit $\ell\rightarrow 0$ of $\mathcal{M}_{kq,\ell}^{ni}$ must be independent of the regularization kernel, thermal effects regularize the dynamics differently. In most cases, one expects $\mathcal{H}_{kq}\neq 0$. Now, I will demonstrate that $\mathcal{Q}_{kq}^{flux}$ will vanish unless there are regions in the system in which the density gradients exhibit singularities.  

 The interscale flux $\mathcal{Q}^{flux}_{kq,\ell}$ is defined in Eqs. \ref{cascadestress} and \ref{interscaleintegrand}. We need to find bounds for $\mathcal{Q}^{flux}_{kq,\ell}$ as a function of the arbitrary coarse-graining length $\ell$. To obtain these bounds, I will closely follow \cite{EyinkDrivas2017} (see also \cite{Eyink_Course}, Chapter IIb ).

   We have from Minkowski's integral and H{\"o}lder generalized inequalities that   
  
 \begin{equation}
 \Vert \mathcal{Q}_{\ell}^{flux} \Vert_{q/3}\leq \frac{1}{2}\int d^dx^{'}\int_{0}^{1}d\lambda\vert V(\mathbf{x}^{'})\vert \Vert \nabla_k\overline{\rho}_{\ell}\Vert_q \Vert \tau_{\ell}(\rho, F_k)(\mathbf{x}_2)\Vert_{q/2}\;\;\;\;\;\; \forall q\geq 3.
 \label{MinkHoldercascadeterm}
 \end{equation}
 
 For the gradient of the coarse-grained density we obtain the following bound 
\begin{equation}
\Vert\nabla\rho(\mathbf{x}+\mathbf{a})\Vert_p\leq \frac{1}{\ell}\int d^d r \vert(\nabla G)_{\ell}(\mathbf{r}-\mathbf{a})\vert\Vert \delta\rho(\mathbf{x};\mathbf{r})\Vert_p=\mathcal{O}\left(\frac{\delta\rho(\ell)}{\ell}\right),
\label{gradientdensityboundw}
\end{equation}
where $\delta\rho(\ell)$ is a typical density increment over the resolution scale $\ell$. Assuming local Besov regularity, we derive for the $L_p$-norm of the coarse-grained density gradients 

\begin{equation}
\frac{1}{\ell}\int d^d r \vert(\nabla G)_{\ell}(\mathbf{r}-\mathbf{a})\vert\Vert \delta\rho(\mathbf{x};\mathbf{r})\Vert_p\leq \frac{1}{\ell}\int d^d r \vert(\nabla G)_{\ell}(\mathbf{r}-\mathbf{a})\vert \vert \mathbf{r} \vert^{\xi_p}.
\label{normcggradientsbesov}
\end{equation}

 Making the change of variable $\varrho=\frac{\mathbf{r}-\mathbf{a}}{\ell}$ we obtain for the last integral 

\begin{equation}
\frac{1}{\ell}\int d^d r \vert(\nabla G)_{\ell}(\mathbf{r}-\mathbf{a})\vert \vert \mathbf{r} \vert^{\xi_p}=\frac{1}{\ell}\int d^d \varrho \nabla G(\varrho)\vert \varrho \ell+a\vert^{\xi_p}\leq\ell^{\xi_p-1}\int d^d \varrho \nabla G(\varrho)\vert\varrho\vert^{\xi_p}.
\label{changevariables}
\end{equation}

To derive the last inequality we applied Theorem 199 of \cite{INEQUALITIES}, since the Besov exponents satisfy $0<\xi_p\leq 1$, and we have for the regularization kernel that $\int d^d \varrho \nabla G(\varrho)=0$.


 From Eq. \ref{identitycumulantsstress} we get for the "unresolved" or "fluctuating" part of the inter-particle forces 
\begin{align}
\Vert \tau_{\ell}(\delta\rho, \delta\mathbf{F})(\mathbf{x}+\mathbf{a})\Vert_{q/2}\leq \int d^dr G_{\ell}(\mathbf{r}-\mathbf{a})\Vert\delta\rho(\mathbf{r};\mathbf{x})\Vert_q\Vert\delta F(\mathbf{r};\mathbf{x})\Vert_q+\nonumber \\
\int d^dr G_{\ell}(\mathbf{r}-\mathbf{a})\Vert\delta F(\mathbf{r};\mathbf{x})\Vert_q\int d^dr^{'} G_{\ell}(\mathbf{r}^{'}-\mathbf{a})\Vert\delta\rho(\mathbf{r}^{'};\mathbf{x})\Vert_q. 
\label{inequalitycumulantsstressq}
\end{align}

Then, from the last inequality we obtain the bound 

\begin{equation}
\Vert \tau_{\ell}(\rho, \mathbf{F})(\mathbf{x}+\mathbf{a})\Vert_{q/2}= \mathcal{O}(\delta\rho(\ell)^2),
\label{inequalitycumulantsfinal}
\end{equation}

 Eqs. (\ref{inequalitycumulantsfinal}), (\ref{gradientdensityboundw}) and (\ref{normcggradientsbesov}) yield the following bound
\begin{equation}
 \Vert \mathcal{Q}_{\ell}^{flux} \Vert_{q/3}=\mathcal{O}\left(\frac{\delta\rho(\ell)^{3}}{\ell}\right)=\mathcal{O}\left(\ell^{3\xi_q-1}\right)\;\;\;\;\;\; \forall q\geq 3.
 \label{cascadefinalestimate}
 \end{equation}

 From the estimate in \ref{cascadefinalestimate} we see that a non-vanishing interscale flux $\mathcal{Q}^{flux}$ would require to have regions in which the Besov exponents of the density field satisfy the inequality
 
  \begin{equation}
  \xi_q\leq 1/3 \;\;\;\; \forall q\geq 3,
 \label{K41scaling}
  \end{equation}
i.e. there must be regions within the material in which the density gradients are singular. 


In typical experiments, the external forcings are applied on the large scales of the systems. Such scenarios imply

\begin{equation}
   \mathcal{F}^{ni}_{kq}+ \mathcal{F}^{e}_{kq}+\mathcal{Q}_{kq}^{thermal}\approx  \mathcal{D}-\lim_{\ell\rightarrow 0}\left(\mathcal{M}_{kq,\ell}^{ni}+\mathcal{N}_{kq,\ell}^e+\mathcal{Q}_{kq,\ell}^{flux}\right).
    \label{RGbalanceapprox}
\end{equation}
 At small load conditions, observations indicate that irreversible stress dissipation occurs at very small scales \cite{Barbot_PhysRevE.97.033001}. Since both \(\mathcal{M}_{kq,\ell}^{ni}\) and \(\mathcal{N}_{kq,\ell}^e\) are large-scale quantities, it is expected that \(\mathcal{Q}_{kq,\ell}^{flux}\) will provide the main contribution to stress dissipation at small scales. There must be regions within the material where density gradients are singular, with the scaling of these density gradients adhering to the inequality \ref{K41scaling}.

 To align with the findings of \cite{Wu2023}, one must show that the inequality \ref{cascadefinalestimate} holds in regions where the eigenvectors of the vibrational modes exhibit a high density of frustrated interfaces with a saddle shape. This will be addressed in a separate publication.

In deriving these results, I did not take an average over different ensembles; rather, the results are applicable to individual realizations. These calculations indicate that the tensors representing the sources and sinks of stress depend on spatial averages of powers of the density increments. This observation is consistent with the fact that the indicator introduced by \cite{Wu2023} is related to a weighted average of modes that possess various characteristic length scales.

  \section{Concluding Remarks}
 In this work, I argued that coarse-grained solutions of the noiseless athermal Dean equation should govern the dynamics of amorphous solids in the limit $T_v\rightarrow 0$. This conclusion stems from the sole hypothesis that plasticity in amorphous solids is a manifestation of an athermal stress dissipative anomaly. From an exact "scale-by-scale stress budget equation," I identified the tensorial quantities responsible for anomalous stress dissipation in amorphous materials, which depend critically on spatial averages of powers of the density increments. This finding aligns with the notion that the indicator introduced in \cite{Wu2023} is related to a weighted average of modes with different characteristic length scales. Thus, the non-perturbative approach developed here allowed the identification of the structural variables linked to sources of plastic deformations from elementary mechanical considerations. One concludes from equation \ref{cascadefinalestimate} that anomalous stress dissipation is not solely due to dislocations; instead, milder singularities in the density field may also contribute to sources of plastic deformations by enabling a nonequilibrium flux of stresses toward arbitrarily small scales. The latter could also be a key mechanism in finite $T_v$ regimes. Consequently, in contrast to standard approaches that focus on Gibbs free-energy minimization, this work suggests that nonequilibrium processes underlie the observed irreversible stress dissipation during small-scale particle rearrangements. The predictions concerning density singularities leading to conservation-law anomalies of mechanical stresses due to nonlinear "cascade" mechanisms can be tested through experiments and numerical simulations.


   Implicit in the derivations of this work is the idea that the system's physics is governed by rare events at very low temperatures. However, it remains uncertain whether a law of large numbers underlies the behavior of amorphous solids at low temperatures. One could address this by directly analyzing the fine-grained stress balance derived from the stochastic Dean equation.

   From first principles, these calculations pave the way for a rigorous characterization of the mechanisms that lead to stress dissipation when the dimensionless number $T_v$ approaches zero. Given the general applicability of these calculations, this formalism can be extended to a wide range of materials, from ductile glasses to ultra-stable glasses that exhibit brittle failure. In light of these results, it would also be interesting to explore whether the nucleation of shear bands in disordered media resembles a Berezinskii-Kosterlitz-Thouless (BKT) phase transition, driven by the excitations of small-scale regions with low Besov regularity.

\section{Acknowledgments}

I am indebted to Reinaldo García-García, who suggested deriving a closed evolution equation for the stress tensor from the stochastic Dean equation. I here did not tackle the closure problem directly; instead, I found a way to bypass it by working in analogy to Onsager's ideal turbulence theory. I also thank Damien Vandembroucq and Sylvain Patinet for their  insights about the phenomenology of plasticity in amorphous solids.

\end{abstract}

\maketitle

\nocite{*}

\bibliography{bible}

\end{document}